\documentclass[traditabstract]{aa} 
\usepackage{graphicx}
\usepackage{txfonts}

\newcommand{\myrule}{\rule[-0.1cm]{0.cm}{0.7cm}} 
\newcommand\lsun{\rm{L_{\odot}}}
\newcommand\msun{\rm{M_{\odot}}}

\newcommand\mjup{\rm{M_{Jup}}}

\begin{document}

\title{Orbit of the young very low-mass spectroscopic binary CHXR\,74
\thanks{Based 
     on observations obtained at the Very Large Telescope of the 
	    European Southern Observatory at Paranal, Chile 
	    with UVES in program 
	    65.I-0011(A),     
	    72.C-0653(A),     
	    75.C-0851(C),     
            77.C-0831(A+D),   
	    380.C-0596(A),    
	    082.C-0023(A),    
	    087.C-0962(B),    
	    and with NACO in program
	    380.C-0596(B).     
	  }}

 \titlerunning{Orbit of the young very low-mass spectroscopic binary CHXR\,74}

   \author{V. Joergens
          \inst{1,2}
          \and
          M. Janson\inst{3}
	  \and
	  A. M\"uller\inst{2}
          }

   \institute{
	     Zentrum f\"ur Astronomie Heidelberg, 
	     Institut f\"ur Theoretische Astrophysik,
	     Albert-Ueberle-Str. 2, 69120 Heidelberg, Germany
	     \and
	     Max-Planck Institut f\"ur Astronomie, 
	     K\"onigstuhl~17, D-69117 Heidelberg, Germany,
             \email{viki@mpia.de}
	 \and
	 Princeton University, Dept. of Astrophysics, 
	 4 Ivy Lane, Princeton, NJ 08544, USA
             }

   \date{Received Oct 5, 2011; accepted Oct 24, 2011}

  \abstract
   {The pre-main sequence star CHXR\,74 (M4.25) in Chamaeleon\,I
     was found a few years ago to be a very low-mass spectroscopic binary.
     A determination of its mass
     would provide a valuable dynamical mass measurement at young ages in the poorly constrained
     mass regime of $<0.3\,\msun$.
     We carried out follow-up radial velocity monitoring with UVES/VLT between 2008 and 2011 and 
     high-resolution adaptive-optic-assisted imaging with NACO/VLT in 2008 with the aim of 
     constraining the binary orbit.
     We present an orbital solution of the system 
     based on the combined radial velocity data set, which spans more than eleven years of 
     UVES monitoring for CHXR\,74.
     The best-fit Kepler model has an orbital
     period of 13.1\,years, zero eccentricity, and a radial velocity semi-amplitude of 2.2\,km\,s$^{-1}$.
     A companion mass $M_2\sin i$ (which is a lower limit due to the unknown orbital inclination $i$)
     of 0.08\,$\msun$ is derived by using a model-dependent mass estimate for the primary of 0.24\,$\msun$.
     The binary separation ($a_1 \sin i$ + $a_2 $) for an inclination of 90$^{\circ}$ is 3.8\,AU, 
     which corresponds to 23\,mas. 
     Complementary NACO/VLT images of CHXR\,74 were taken
     with the aim to directly resolve the binary. 
     While there are marginal signs of an extended point spread function (PSF),
     we have detected no convincing companion to CHXR\,74 in the NACO images.
     From the non-detection of the companion together with a prediction of the
     binary separation at the time of the NACO observations, we derive an upper limit
     for the $K$-band brightness ratio of the two binary components of 0.5.
     This allows us to estimate an upper limit of the companion mass of 0.14\,$\msun$ 
     by applying evolutionary models. 
     Thus, we confirm that CHXR\,74 is a very low-mass spectroscopic binary
     and constrain the secondary mass to lie within the range of about 0.08 and 0.14\,$\msun$. 
     We predict an astrometric signal of the primary between 0.2 and 0.4\,mas when taking into account 
     the luminosity of the companion.
     The GAIA astrometric mission might well be able to solve the astrometric orbit of the primary
     and in combination with the presented radial velocity data determine an absolute companion mass.
}


   \maketitle
%

\section{Introduction}
\label{sect:intro}

Binaries whose orbital periods are sufficiently short for us to follow their orbital motion 
in a reasonable time are key astronomical objects because they allow us dynamical mass 
measurements. Masses are the very foundation of our understanding of star-, brown dwarf- (BD) and 
planet formation and evolution, e.g. they define the IMF and are the most important 
input parameter for evolutionary models. 
A means to search for companions with orbital distances of a few AU are 
spectroscopic surveys that monitor the radial velocity (RV) shift caused by a close companion.  
From the spectroscopic orbit alone, only lower mass limits $M\sin i$
can be determined because of the unknown inclination $i$. Spectroscopic binaries for which the
astrometric or visual orbit can be determined or that happen to be eclipsing systems allow absolute mass
determinations.
In the mass regime $<$0.3\,M$_\mathrm{\odot}$, a few spectroscopic binaries
have been detected and their orbits determined 
(e.g., Basri \& Mart\'\i n 1999; Stassun, Mathieu \& Valenti 2006;
Joergens \& M\"uller 2007; Joergens et al. 2010; Blake et al. 2008).
Furthermore, follow-up RV measurements permitted 
additional constraints of the orbits of
a few close visual BD and very low-mass (VLM) binaries (e.g., Zapatero Osorio et al. 2004; Simon, Bender \& Prato 2006;
Konopacky et al. 2010).
For masses $<$0.3\,M$_\mathrm{\odot}$, evolutionary models rely only on very few mass 
measurements for very young binaries (Mathieu et al. 2007; Stassun et al. 2006).
While there used to be little or no overlap between spectroscopic, and, consequently close
binaries and visual binaries, particularly at the distance of star-forming regions,
the situation is changing owing to
recent advances in spatial resolution achieved by
adaptive optics (AO) and interferometric instruments. 

Within the course of an RV survey for binaries among
very young BDs and low-mass stars with UVES/VLT in the Chamaeleon\,I (Cha\,I) star-forming cloud,
the very low-mass star CHXR\,74 (M4.25, $\sim$0.24\,$\msun$) was detected 
to be a spectroscopic binary with a presumably relatively long orbital period (Joergens 2006, 2008).

We have carried out follow-up UVES observations between 2008 and 2011 to constrain the
spectroscopic orbit.
Furthermore, we have obtained high-resolution AO images with NACO/VLT of CHXR\,74 
with the aim to directly resolve the binary.
We present here the results of these observations, which include the determination of an RV orbit solution
and of an upper limit for the brightness ratio.
Finally, we derive constraints for the masses of the two binary components
and investigate prospects for future dynamical mass measurements.


\section{CHXR\,74}
\label{sect:chxr74}

\object{CHXR\,74} was identified as low-mass member of the Cha\,I
star-forming cloud with spectral type M4.5 by Comer\'on et al. (1999). 
CHXR\,74 has a brightness of 17.3\,mag in $V$ band and 10.2\,mag in $K_{\rm S}$ band 
(Comer\'on et al. 1999; Luhman 2004) and is classified as class III object based on
its spectral energy distribution,
i.e. it has no detected disk (Luhman et al. 2008). 
The most recent determination of stellar parameters for CHXR\,74 yields  
a spectral type of M4.25, an effective temperature of $T_{\rm eff}$=3234\,K, 
and a bolometric luminosity of $L=0.15\,\lsun$ (Luhman 2007).
Using these stellar parameters of the \emph{unresolved system},
we estimate a mass of about 0.25\,$\msun$ and an age of about 2\,Myr
for CHXR\,74 based on evolutionary model tracks (Baraffe et al. 1998).
We show below that the companion contributes significantly to the total luminosity
of the system and that the model-dependent mass of the primary is
0.23-0.24\,$\msun$, instead 
(cf. Sect.\,\ref{sect:concl}).

\begin{table}
\begin{minipage}[t]{\columnwidth}
\begin{center}
\caption{
\label{tab:rvs} 
RV measurements of CHXR74.
}
\renewcommand{\footnoterule}{}  
\begin{tabular}{llll}
\hline
\hline
\myrule
Date      & HJD           & RV     & ~$\sigma_{RV}$\\
          &               & [km\,s$^{-1}$] & [km\,s$^{-1}$] \\
\hline
\myrule
2000 03 13 &  2451616.787145 & 15.829 \tablefootmark{a} & 0.012  \\ 
2000 03 31 &  2451634.520915 & 15.211 \tablefootmark{a} & 0.216  \\ 
2000 04 22 &  2451656.512470 & 15.770 \tablefootmark{a} & 0.095  \\ 
2000 05 21 &  2451686.482610 & 14.822 \tablefootmark{a} & 0.031  \\ 
2004 03 03 &  2453067.829560 & 17.863 \tablefootmark{a} & 0.033  \\ 
2004 03 12 &  2453076.664875 & 18.223 \tablefootmark{a} & 0.005  \\ 
2004 03 21 &  2453085.770220 & 18.282 \tablefootmark{a} & 0.109  \\ 
2004 03 24 &  2453088.798215 & 18.020 \tablefootmark{a} & 0.083  \\ 
2004 03 31 &  2453095.780420 & 19.003 \tablefootmark{a} & 0.448  \\ 
2004 04 01 &  2453096.776000 & 18.779 \tablefootmark{a} & 0.008  \\ 
2005 03 21 &  2453450.589225 & 19.468 \tablefootmark{a} & 0.172  \\ 
2006 04 10 &  2453835.634560 & 18.845 &     \\ 
2006 06 15 &  2453901.522670 & 19.642 &     \\ 
\hline
\myrule
2008 02 19 &  2454515.798960 & 18.712 &    \\ 
2008 03 07 &  2454532.786240 & 19.202 &    \\ 
2008 03 19 &  2454544.569660 & 19.719 &    \\ 
2009 01 12 &  2454843.867690 & 18.700 &    \\ 
2009 01 14 &  2454845.857240 & 18.595 &    \\ 
2009 01 16 &  2454847.754840 & 18.628 &    \\ 
2011 04 06 &  2455657.749380  & 16.396 & \\ 
2011 05 27 &  2455708.561260  & 16.168 & \\ 
2011 07 03 &  2455746.490790  & 15.414 & \\ 
\hline
\end{tabular}
\tablefoot{Listed are new RV data (bottom panel) and 
previous RV measurements (top panel), which were re-processed for this work.
HJD is given at the middle of the exposure; 
$\sigma_{RV}$ is the sample standard deviation of two consecutive measurements.
The error of the relative RVs taking into account RV noise caused by activity is estimated to
about 450\,m\,s$^{-1}$. 
An additional error of about 400\,m\,s$^{-1}$ has to be taken into account for the absolute RVs.}
\tablefoottext{a}{RV value is the average of two single consecutive measurements.}
\end{center}
\end{minipage}
\end{table}


\section{Radial velocities and orbital solution}
\label{sect:rvs}

\begin{figure}[t]
\centering
\includegraphics[width=\linewidth,clip]{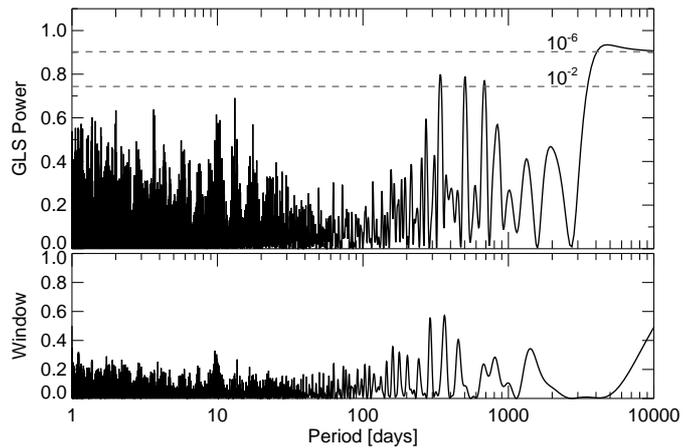}
\caption{
\label{fig:period}
GLS periodogram (upper plot) of the RV data, and the corresponding window function (lower plot). 
The two horizontal dashed lines indicate FAP levels of $10^{-2}$ and $10^{-6}$. 
The highest peak at a period of 4785\,days has an FAP of $2.3\cdot10^{-8}$.
}
\end{figure}

\begin{figure}[t]
\centering
\includegraphics[width=\linewidth,clip]{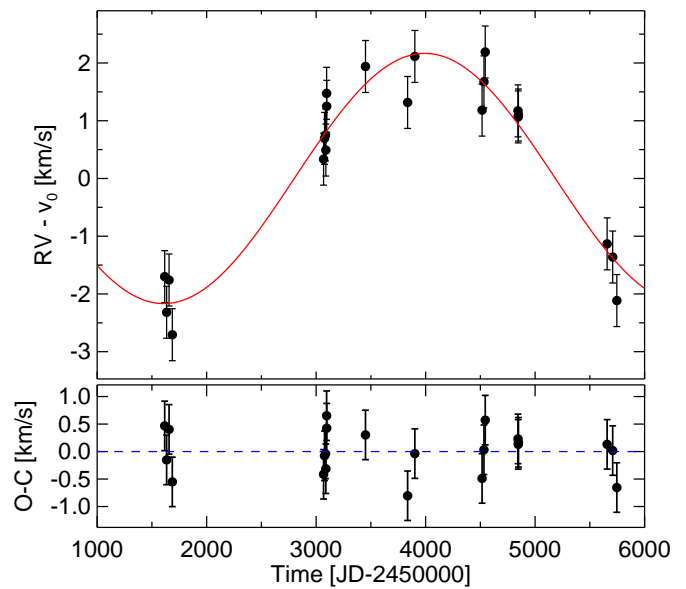} 
\caption{
\label{fig:orbit}
Upper panel: RV measurements of CHXR\,74 between 2000 and 2011 based on UVES/VLT spectra. 
The system velocity $V_0$ was subtracted.
The red solid line shows the best-fit Keplerian orbit for a period of 4770\,days (13.1\,years)
and 
a semi-amplitude of 2.2\,km\,s$^{-1}$. Lower panel: RV residuals.
}
\end{figure}

Spectroscopic observations of CHXR\,74 were carried out between 2000 and 2011 with
the \emph{Ultraviolet and Visual Echelle Spectrograph} (UVES, Dekker et al. 2000) 
attached to the VLT 8.2\,m KUEYEN telescope
at a spectral resolution $\lambda$/$\Delta \lambda$ of 40\,000 in the 
red optical wavelength regime.
The RVs were measured from these spectra based on a cross-correlation technique
employing telluric lines for the wavelength calibration. 
Details of the data analysis can be found in Joergens (2006, 2008).

The RV measurements for spectra taken between 2000 and 2006 
provided evidence that CHXR\,74 has a long-period spectroscopic companion (Joergens 2006, 2008).
Here, we present new RV data based on UVES spectra taken between 2008 and 2011. 
Table\,\ref{tab:rvs} lists the new RV measurements of CHXR\,74
and previous RV data. The latter were reprocessed for this work
by using a slightly refined wavelength range for the cross-correlation to exclude
a region that had been found to be contaminated by telluric lines 
and imperfectly corrected CCD cosmetic blemishes and that introduced additional scatter.
The given errors $\sigma_{RV}$ are the sample standard deviation of 
two consecutive individual measurements.
The RV noise caused by activity is estimated to about 450\,m\,s$^{-1}$ based on 
the night-to-night rms scatter in the data.
This error value is used for finding the RV orbit solution. 

The combined data set spans more than 11 years of RV monitoring for CHXR\,74 between 2000 and 2011
(Fig.\,\ref{fig:orbit}). 
Starting in 2000, we witnessed first an increase of the RV until 2006 
by about 4\,km\,s$^{-1}$ followed by a period of only small RV changes and 
finally a decrease of the RV in 2011 back to the original RV level. 

To identify periodicities in the RV data, we computed the generalized Lomb-Scargle (GLS) periodogram 
(Zechmeister \& K\"urster 2009) and its window function (Fig.\,\ref{fig:period}). 
For a better visibility of longer periods, the GLS periodogram is plotted with a logarithmic scale for the period.

The GLS periodogram reveals a significant peak at a period of 4785\,d with a false-alarm probability (FAP) 
of $2.3\cdot10^{-8}$. In addition, there are several peaks at shorter periods but they have FAPs 
higher than $10^{-3}$ and are not considered to be significant. These peaks are expected to be aliases because
the periodogram is a convolution of a present signal in the data and the sampling function of these data (window function). 
For example, the alias peak in the periodogram at 340\,d is caused by the 365\,d sampling period 
(yearly alias, highest peak in the window function).

We calculated an orbital solution based on the RV data  
using $\chi^2$ minimization.
We fitted the four free parameters period $P$, 
periastron time $T$, RV semi-amplitude $K$, and system velocity $V_{0}$. 
The eccentricity $e$ was fixed to zero in this fit because treating it as free parameter
leads to very high eccentricity values, poorly constrained longitudes of periastron,
and unreasonable values for the lower limit of the companion mass $M_2 \sin i$ (several times higher than $M_{1}$). 
We note that such unrealistic results occur for $e>0.2$, while derived orbit
parameters for $e \leq 0.2$ are consistent within the errors with the values derived for $e=0$.

The best-fit Kepler orbit is shown in Fig.\,\ref{fig:orbit}.
It has a period of 4770\,days (13.1\,yr) and an RV semi-amplitude of 2.2\,km\,s$^{-1}$. 
The derived lower limit to the semi-major axis of the primary $a_1 \sin i$ is 0.95\,AU.
The reduced $\chi^2$ of the orbital fit is 0.94.
The complete list of determined orbital elements is given in Table\,\ref{tab:orbitparam}.

The minimum mass $M_2 \sin i$ and the semi-major axis $a_2$ of the secondary 
for a single-lined RV orbit depend on the primary mass.
The mass estimate for CHXR\,74 based on the luminosity and effective temperature 
of the unresolved source by applying evolutionary models is 0.25\,$\msun$ (cf. Sect.\,\ref{sect:chxr74}).
We show in Sect.\,\ref{sect:concl} 
that the companion contributes significantly to the total luminosity of the system, which
leads to a reduced primary mass.
Using $M_1$=0.24\,$\msun$, the mass $M_2 \sin i$ of the companion is
determined to be 84\,$\mjup$ (0.08\,$\msun$) and its semi-major axis $a_2$ to be 2.84\,AU.
The errors in $M_2 \sin i$ and $a_2$ given in Table\,\ref{tab:orbitparam}
are based on the fit and do
not take into account additional possible errors in the primary mass,
such as those introduced by the evolutionary models used to estimate the primary mass.


\begin{table}[t]
\begin{center}
\caption{
\label{tab:orbitparam} 
Orbital and physical parameters derived for the best-fit Keplerian model
of CHXR\,74.
}
\begin{tabular}{lc}
\hline
\hline
\\

Parameter  & Value \\

\hline
\myrule

$P$ (days)                \dotfill  & 4770 $\pm$  386    \\
$T$ (HJD-2450000)         \dotfill  & 1608 $\pm$  204    \\
$e$                       \dotfill  & 0.0 (fixed)    \\
$\omega$ ($^{\circ}$)     \dotfill  & 0.0 (fixed)    \\
$K$ (km\,s$^{-1}$)        \dotfill  & 2.17 $\pm$  0.14    \\
$V_0$ (km\,s$^{-1}$)      \dotfill  & 17.528 $\pm$ 0.17   \\
\hline
$f(m)$ ($10^{-3}\,\msun$) \dotfill  & 5.026  $\pm$  1.061  \\
$M_2 \sin i $ ($\mjup$)   \dotfill  & 84   $\pm$ 6 \\
$M_2 \sin i $ ($\msun$)   \dotfill  & 0.08 $\pm$ 0.006 \\
$a_1 \sin i$ (AU)         \dotfill  & 0.95 $\pm$ 0.08 \\
$a_2 $ (AU)               \dotfill  & 2.84 $\pm$ 0.20 \\

\hline

$N_{\rm meas}$           \dotfill   & 22    \\
Span (days)              \dotfill   & 4130  \\ 
$\sigma$ (O-C) (m/s)     \dotfill   & 395   \\
$\chi^{2}_{\rm red}$     \dotfill   & 0.94  \\

\hline
\end{tabular}
\tablefoot{
The given parameters are: orbital period,  
periastron time, eccentricity, longitude of periastron,
RV semi-amplitude, system velocity, 
mass function, lower limit of the companion mass,
lower limit of the semi-major axis of the primary,	  
semi-major axis of the companion, number of measurements,
time span of the observations, residuals, reduced $\chi^2$. 
A model-dependent primary mass $M_1$=0.24\,$\msun$ was used 
to derive $M_2 \sin i $ and $a_2 $; the given
errors of $M_2 \sin i $ and $a_2 $ are solely based on the fit.
See text for more details.
}
\end{center}
\end{table}

\section{NACO observations}
\label{sect:naco}

A binary separation of about 3\,AU or more was predicted for CHXR\,74 based on RV data
from 2000 to 2006 (Joergens 2008). This corresponds to an angular separation of almost 20 milli\,arcseconds (mas) or more
at the distance of Cha\,I (160-165\,pc). 
Therefore, it seemed to be within the realms of possibility to directly resolve this very young spectroscopic binary system 
with current direct imaging facilities.
A resolved orbit would yield the orbital inclination and with it an absolute companion mass
and absolute semi-major axes
as well as an independent measurement of the distance.

For this purpose, we carried out adaptive optics (AO) imaging of CHXR\,74 with \emph{NAOS Conica} 
(NACO, Lenzen et al. 2003; Rousset et al. 2003)
attached to the VLT 8.2\,m YEPUN telescope, on 16 Jan 2008. 
Observations were made with IR wavefront sensing (N90C10 dichroic) in $K$-band with the S13 camera providing a pixel scale of 13.3\,mas/pixel, and with the target itself as natural guide star (NGS). Data were taken at two instrument rotator angles separated by 90$^{\rm o}$. At each angle, 11 frames were taken with jittering applied. Each frame consisted of a single 12\,s exposure, giving a total integration time of 132\,s per angle. The data were reduced and collapsed into a final 0$^{\rm o}$ and a final 90$^{\rm o}$ frame using the 'jitter' routine of the ESO 'eclipse' package.

There is no convincing companion detected to CHXR\,74 in the images. The point spread function (PSF) 
has a FWHM of about 70\,mas and
is slightly extended in the north/south direction (along close to the y-axis in the 0$^{\rm o}$ image and to the x-axis in the 90$^{\rm o}$ image). The extension direction is clearly distinct from the parallactic angle during the observations ($\sim$-40$^{\rm o}$), which excludes the possibility that it is a result of differential atmospheric refraction. Hence, it is in principle possible that this extension is caused by the companion. However, because the signal is rather small, and because 
some instrumental effects are hard to exclude (e.g., an unfortunate PSF variation between the 0$^{\rm o}$ and 90$^{\rm o}$ frames), we do not consider this as a reliable detection. Instead, we used the data to quantify upper limits to the brightness ratio and lower limits to the projected separation by excluding parameters that would cause a stronger impact on the PSF than what we observe, 
to a statistically significant degree.

This was performed by introducing fake companions in the data at different brightness ratios and separations, 
and evaluating the signal-to-noise ratio ($S/N$) of the resulting signature. The effect on the PSF as a function of companion properties is quantified in terms of the PSF width along the extended direction, compared to the width along the perpendicular direction. The procedure is the following: The 0$^{\rm o}$ and 90$^{\rm o}$ pair of images of CHXR\,74 with no artificial companions are taken as the reference cases of a star with no (adequately significant) companion signature. For this pair of images, we loop through position angles of 0$^{\rm o}$ through 170$^{\rm o}$, in steps of 10$^{\rm o}$. For each angle, a cross-section of the PSF at that angle is sampled through bilinear interpolation along the cut. A Gaussian fit is performed on each cross-section, evaluating in particular the FWHM. The array of FWHMs at 0$^{\rm o}$ and 90$^{\rm o}$ are denoted $w_{\rm ref, 0}$ and $w_{\rm 
 ref, 90}$, respectively. These are subtracted pairwise to create $\Delta w_{\rm ref} = w_{\rm ref, 0} - w_{\rm ref, 90}$. 
This means that if the PSF were significantly extended along 50$^{\rm o}$, $\Delta w$ would show a sinusoidal-like variation, with a maximum at 50$^{\rm o}$ and a minimum at 140$^{\rm o}$. As we discussed above, $\Delta w_{\rm ref}$ does weakly display such a trend with a peak at close to 0$^{\rm o}$. The total scatter (i.e., the standard deviation) of $\Delta w_{\rm ref}$ is taken as the noise level $N$.

\begin{figure}[t]
\centering
\includegraphics[width=\linewidth,clip]{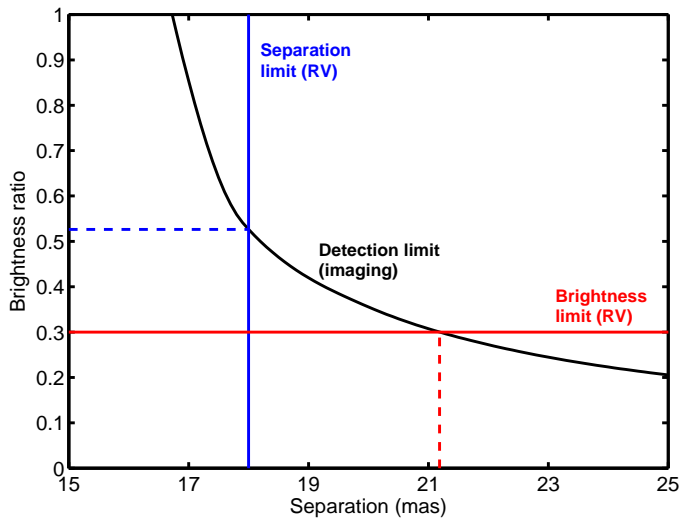} 
\caption{
\label{fig:naco}
Detection map of the NACO observations of CHXR\,74. 
Displayed are the parameter space of detectable companions
in terms of $K$-band brightness ratio and binary separation.
The black curve denotes the detection limit for a S/N of 5.
The blue vertical line represents the minimum binary separation
(18\,mas) at the time of the NACO observations and
the red horizontal line the model-dependent lower limit 
of the $K$-band brightness ratio, both derived from the RV orbit.
See text for more details.}
\end{figure}

We then systematically introduced artificial companions in the images. These are copies of the primary PSF, added at different separations and with different brightness ratios. A grid was used for the sampling that sampled the separation in steps of 0.15 pixels, from 0.15 pixels to 3.00 pixels, and the brightness ratio in steps of 0.05, from 0.05 to 1.00. The $\Delta w$ corresponding to a simulated case is denoted $\Delta w_{\rm sim}$. For each case we evaluated the signal $S$ as the rms of the $\Delta w_{\rm sim} - \Delta w_{\rm ref}$ components. We set $S/N > 5$ as the criterion for detectability. 
In Fig.\,\ref{fig:naco} we show the parameter ranges that remain allowed and can be excluded, 
respectively, on the basis of the images.

\section{Conclusions}
\label{sect:concl}

We confirm that CHXR\,74 (M4.25) is a very low-mass pre-main sequence spectroscopic binary
based on new RV measurements obtained between 2008 and 2011 with UVES at the VLT.
The Kepler orbit that fits the combined RV data set (2000-2011) best  
has an orbital period of 13.1\,years and zero eccentricity. 
A companion minimum mass $M_2\sin i$ of 0.08\,$\msun$ was derived by using a model-dependent 
mass estimate for the primary of 0.24\,$\msun$.
The binary separation ($a_1 \sin i$ + $a_2 $) for an inclination of 90$^{\circ}$ is 3.8\,AU, 
which corresponds to 23\,mas at the distance of Cha\,I (160-165\,pc). 
Complementing the RV measurements,
high-resolution AO images of CHXR\,74 were taken with NACO at the VLT 
with the aim of directly resolving the binary. 
While there are marginal signs of an extended PSF,
we have no convincing companion detected to CHXR 74 in these images.

In the following, we investigate the parameter space of binary properties that is
allowed for CHXR\,74 given the presented UVES and NACO observations.
First, we calculate the absolute companion mass $M_2$ for different inclinations
based on the RV orbit and taking the companion's luminosity into account. 
For this calculation, $M_2$ was derived directly from the mass function $f(m)$ rather than
using the value of $M_2 \sin i$ 
given in Table\,\ref{tab:orbitparam}, which is for non-negligible companion masses 
valid only for large inclination angles.
The luminosity ratio $L_2 / L_1$ and the primary mass $M_1$ were then estimated for different inclinations
by employing low-mass evolutionary model tracks (Baraffe et al. 1998) and assuming the coevality of the
binary components. 
For the luminosity of the unresolved system and its effective temperature, 
the values determined by Luhman (2007, cf. Sect.\,\ref{sect:chxr74}) were used.
While they correspond to a mass estimate of 0.25\,$\msun$, 
the fact that the secondary of CHXR\,74 contributes significantly to the total luminosity
leads to a lower primary mass estimate depending on the inclination.

At minimum, i.e. for an inclination $i$=90$^{\circ}$, the mass of the companion is estimated to be 0.08\,$\msun$. 
In this case the luminosity ratio $L_2 / L_1$ is 0.30 and
the primary has a mass of about 0.24\,$\msun$.
For smaller inclination angles, the mass and luminosity of the companion increases, e.g.,  
for $i$=60$^{\circ}$, a companion mass of 0.10\,$\msun$ is derived.
In this case, $L_2 / L_1$ is estimated to 0.36 and $M_1$ to about 0.24\,$\msun$. 
The limiting case is to consider a companion that is as luminous as the primary
(i.e., $L_1$=$L_2$=$L_{\rm{tot}}/2$), although we show in the next paragraph
that this is unrealistic. 
Comparison with evolutionary tracks shows that then
both components would have a mass of $M_1=M_2\approx 0.23\,\msun$. 
This implies a minimum orbital inclination of 27$^{\circ}$, as derived from the mass function.

We show in this paragraph that the range of possible inclinations and companion masses
constrained by the RV data can be further restricted based on the NACO observations.
The binary separation of CHXR\,74 on the date of our NACO observations is predicted 
to be 3.0\,AU at minimum ($i$=90$^{\circ}$) based on the RV orbit solution
(S. Reffert, pers. comm.). 
This corresponds to an angular separation of 18\,mas at the distance of Cha\,I.
Considering the non-detection in the NACO observations and
the derived detectable $K$-band brightness ratio as function of the actual
separation (black curve in Fig.\,\ref{fig:naco}),
the minimum separation of 18\,mas (blue vertical line in Fig.\,\ref{fig:naco}) can be translated into
an upper limit for the $K$-band brightness ratio of the two components in CHXR\,74 of 0.53. 
This brightness ratio is applied to
divide the $K_{\rm S}$-band magnitude measured for the unresolved system
($m_{\rm{K}}$=10.21\,mag) 
among the two components ($m_{\rm{K,1}} \leq$ 10.67\,mag, $m_{\rm{K,2}} \geq$ 11.36\,mag).
Comparing again with evolutionary tracks yields an upper limit for the 
companion mass of 0.14\,$\msun$ and a lower limit for the primary mass
of 0.23\,$\msun$. 
This requires the orbital inclination to be $\geq 40^{\circ}$
(again derived directly from the mass function rather than using $M_2 \sin i$). 
Therefore we conclude from the combined UVES and NACO observations that the secondary 
of CHXR\,74 has a mass within the range of about 0.08 and 0.14\,$\msun$. 

We note that the combined RV and imaging data allow us also to derive 
an upper limit for the binary separation at the time of the
NACO observation of slightly above 21\,mas (3.4\,AU). 
This follows from the minimum mass $M_2 \sin i$ from the RV orbit (0.08\,$\msun$),
which can be converted into a model-dependent 
lower limit for the $K$-band brightness ratio of about 0.3, which in turn
can be translated into a separation via the NACO detection map (red horizontal line in Fig.\,\ref{fig:naco}).
This means, on one hand, that we have apparently just missed the companion in the NACO observations.
On the other hand, it demonstrates that the orbital period cannot be dramatically longer
than the value determined from the RV data (13.1\,yr), which sample about 87\% of the orbit.

The astrometric space mission GAIA
might be able to monitor the astrometric orbit of the primary in CHXR\,74.
While the minimum semi-major axis of the primary is 0.95\,AU (6\,mas),
the astrometric signature is smaller for
a non-negligible luminosity of the companion
because the astrometric orbit corresponds to the orbit of the photocenter around
the center of mass. 
Taking this into account, we predict an astrometric signature of the primary of CHXR\,74
between 0.4 and 0.2\,mas
for the inclination range between 90$^{\circ}$ and 40$^{\circ}$.
If the astrometric orbit can be determined, it will yield 
in combination with the presented RV orbit solution, 
the absolute mass of the companion and an independent distance measurement.
This would provide a valuable dynamical mass measurement at young ages in the poorly constrained
mass regime of $<$0.3\,$\msun$.

\begin{acknowledgements}
We acknowledge the excellent work of the ESO staff at Paranal, 
who took the data presented here in service mode. 
We thank S. Reffert for calculating the separation prediction
at the time of the NACO observations.
Furthermore, we are grateful for comments from
the referee M. Simon and the editor T. Forveille
that helped to improve the paper.
Part of this work was funded by the ESF in Baden-W\"urttemberg.
\end{acknowledgements}

\listofobjects


\begin{thebibliography}{}
\bibitem{BCAH} Baraffe, I., Chabrier, G., Allard, F. \& Hauschildt, P. H. 1998, A\&A, 337, 403

\bibitem{basri1999} Basri, G., \& Mart\'{\i}n, E. L. 1999, ApJ, 118, 2460 

\bibitem{blake2008} Blake, C. H., Charbonneau, D., White, R. J., Torres, G., Marley, M. S., Saumon~D. 2008
                    ApJ, 678, L125 

\bibitem{comeron99} Comer\'on, F., Rieke, G. H., \& Neuh\"auser, R. 1999, A\&A, 343, 477 

\bibitem{dekker} Dekker, H., D'Odorico, S., Kaufer, A., Delabre, B., \& Kotzlowski, H. 2000, 
                 in SPIE Vol. 4008, ed. by M.Iye, A.Moorwood, 534

\bibitem{J06a} Joergens, V. 2006, A\&A, 446, 1165

\bibitem{JM07} Joergens, V., \& M\"uller, A. 2007, ApJ, 666, L113

\bibitem{J08} Joergens, V. 2008, A\&A, 492, 545

\bibitem{J2010} Joergens, V., M\"uller, A., Reffert, S. 2010, A\&A, 521, A24

\bibitem{konopacky2010} Konopacky et al. 2010, ApJ, 711, 1087

\bibitem{lenzen} Lenzen, R., Hartung, M., Brandner, W. et al. 2003, SPIE 4841, 944 

\bibitem{luhman2004} Luhman, K. L. 2004, ApJ, 614, 398

\bibitem{luhman2007} Luhman, K. L. 2007, ApJS, 173, 104

\bibitem{luhmanetal2008} Luhman, K. L., Allen, L. E., Allen, P. R. et al.
                     2008, ApJ, 675, 1375

\bibitem{mathieu2007} Mathieu, R. D., Baraffe, I., Simon, M., Stassun, K. G. \& White, R., 2007,
in Protostars and Planets\,V, ed. by B. Reipurth, D. Jewitt \& K. Keil,
(Tucson, University of Arizona Press), 411

\bibitem{rousset} Rousset, G., Lacombe, F., Puget, P. et al. 2003, SPIE 4839, 140

\bibitem{simon2006} Simon, M., Bender, C. \& Prato, L. 2006, ApJ, 644, 1183

\bibitem{stassun2006} Stassun, K. G., Mathieu, R. D. \& Valenti, J. A. 2006, Nature, 440, 311

\bibitem{zapatero2004} Zapatero Osorio, M. R., Lane, B. F., Pavlenko, Ya., 
  Mart\'\i n E. L.,  Britton, M. \& Kulkarni, S. R. 2004, ApJ, 615, 958

\bibitem{zechmeister} Zechmeister, M. \& {K{\"u}rster}, M. 2009, A\&A, 496, 577 

\end{thebibliography}
\end{document}